\documentclass[preprint,aip,jcp]{revtex4-1}
\usepackage{graphicx}
\usepackage{amssymb}
\usepackage{amsmath}
\usepackage{ifthen}
\usepackage{bbm}
\usepackage{xspace}
\usepackage[note-name=, use-sort-key = false]{notes2bib}
\usepackage{longtable}
\usepackage[utf8]{inputenc}
\usepackage[T1]{fontenc}

\newcommand{\denominator}[2]{\tens{\Delta}{#1}{#2}}

\newcommand{\dsrg}[0]{DSRG(2)\xspace}
\newcommand{\dsrgph}[0]{DSRG(2*)\xspace}

\newcommand{\Eh}[0]{$E_{\rm h}$\xspace}
\newcommand{\mref}[0]{\Phi}

\newcommand{\tens}[3]{{#1}_{#2}^{#3}}
\newcommand{\cop}[1]{\hat{a}^{\dagger}_{#1}}
\newcommand{\aop}[1]{\hat{a}_{#1}}
\newcommand{\sqop}[2]{\hat{a}_{#2}^{#1}}

\newcommand{\aphystei}[2]{\langle{#1}||{#2}\rangle}

\newcommand{\no}[1]{ \{ {#1} \}}

\newcommand{\bra}[2][0]
{\ifthenelse{\equal{#1}{0}}{\left\langle #2 \right|}
{\ifthenelse{\equal{#1}{1}}{\big\langle #2 \big|}
{\ifthenelse{\equal{#1}{2}}{\Big\langle #2 \Big|}
{\ifthenelse{\equal{#1}{3}}{\bigg\langle #2 \bigg|}
{\ifthenelse{\equal{#1}{4}}{\Bigg\langle #2 \Bigg|}
{Error}}}}}
}

\newcommand{\bracket}[4][0]
{\ifthenelse{\equal{#1}{0}}{\left\langle #2 \middle| #3 \middle| #4 \right\rangle}
{\ifthenelse{\equal{#1}{1}}{\big\langle #2 \big| #3 \big| #4 \big\rangle}
{\ifthenelse{\equal{#1}{2}}{\Big\langle #2 \Big| #3 \Big| #4 \Big\rangle}
{\ifthenelse{\equal{#1}{3}}{\bigg\langle #2 \bigg| #3 \bigg| #4 \bigg\rangle}
{\ifthenelse{\equal{#1}{4}}{\Bigg\langle #2 \Bigg| #3 \Bigg| #4 \Bigg\rangle}
{Error}}}}}
}

\newcommand{\braket}[3][0]
{\ifthenelse{\equal{#1}{0}}{\left\langle #2 \middle| #3 \right\rangle}
{\ifthenelse{\equal{#1}{1}}{\big\langle #2 \big| #3 \big\rangle}
{\ifthenelse{\equal{#1}{2}}{\Big\langle #2 \Big| #3 \Big\rangle}
{\ifthenelse{\equal{#1}{3}}{\bigg\langle #2 \bigg| #3 \bigg\rangle}
{\ifthenelse{\equal{#1}{4}}{\Bigg\langle #2 \Bigg| #3 \Bigg\rangle}
{Error}}}}}
}

\newcommand{\ket}[2][0]
{\ifthenelse{\equal{#1}{0}}{\left| #2 \right\rangle}
{\ifthenelse{\equal{#1}{1}}{\big| #2 \big\rangle}
{\ifthenelse{\equal{#1}{2}}{\Big| #2 \Big\rangle}
{\ifthenelse{\equal{#1}{3}}{\bigg| #2 \bigg\rangle}
{\ifthenelse{\equal{#1}{4}}{\Bigg| #2 \Bigg\rangle}
{Error}}}}}
}

\begin{document}
%
%
\title{A driven similarity renormalization group approach to quantum many-body problems} 

%
%
\author{Francesco A. Evangelista}
\email{francesco.evangelista@emory.edu}
\affiliation{Department of Chemistry and Cherry L. Emerson Center for Scientific Computation, Emory University, Atlanta, Georgia 30322, USA}

%
%
\date{\today}

%
%
\begin{abstract}
Applications of the similarity renormalization group (SRG) approach [F. Wegner,  Ann. Phys. \textbf{506}, 77 (1994), 
S. D. G{\l}azek and K. G. Wilson, Phys. Rev. D \textbf{49}, 4214 (1994)]
to the formulation of useful many-body theories of electron correlation are considered.
In addition to presenting a production-level implementation of the SRG based on a single-reference formalism, a novel integral version of the SRG is reported, in which the flow of the Hamiltonian is driven by a source operator.  It is shown that this driven SRG (DSRG) produces a Hamiltonian flow that is analogous to that of the SRG.
Compared to the SRG, which requires propagating a set of ordinary differential equations, the DSRG is computationally advantageous since it consists of a set of polynomial equations.
The equilibrium distances, harmonic vibrational frequencies, and vibrational anharmonicities of a series of diatomic molecules computed with the SRG and DSRG approximated with one- and two-body normal ordered operators are in good agreement with benchmark values from coupled cluster with singles, doubles, and perturbative triples [CCSD(T)].
Particularly surprising results are found when the SRG and DSRG methods are applied to C$_2$ and F$_2$.  In the former case both methods fail to converge, while in the latter case an unbound potential energy curve is obtained.
A modified commutator approximation is shown to correct these problems in the case of the DSRG method.
\end{abstract}

%
%
\maketitle
\section{Introduction}
\begin{figure}[b]
\centering
\includegraphics[width=3.375in]{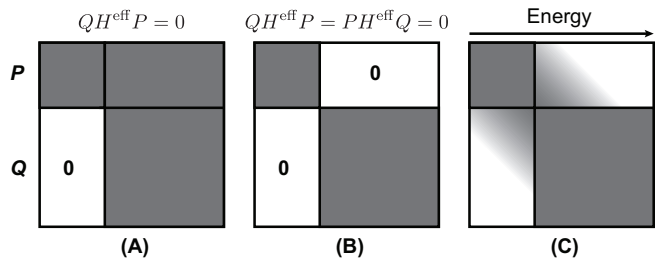}
\caption{Matrix representation of the similarity-transformed Hamiltonian.  (A) and (B) illustrate the decoupling of the $P$ and $Q$ spaces in coupled cluster theory and its unitary variant, respectively.  (C) Decoupling of states with large energy difference achieved by the similarity renormalization group approach.} 
\label{fig:hamiltonian_decoupling}
\end{figure}

A large number of theoretical approaches used to solve the electronic Schr\"odinger equation fall under the broad category of effective Hamiltonian theories.\cite{Bloch:1958wc,Freed:1974vo,Brandow:1977bk}
These include L\"owdin's partitioning method,\cite{Lowdin:1963hj} perturbation theory, coupled cluster (CC) theory\cite{Crawford:2000ub,Bartlett:2007kv,Shavitt:2009uo} and its Fock-space\cite{Mukherjee:1975wy,Lindgren:1978tt,Kutzelnigg:1982wv,Haque:1984vm,Stolarczyk:1985to} and equation-of-motion extensions,\cite{Offermann:1976tg,Stanton:1993be,Nooijen:1995tr,Nooijen:1997kc,Krylov:2001kg,Levchenko:2004vg,Levchenko:2005bf,Krylov:2008ud} as well as a variety of multireference perturbation theories,\cite{Nitzsche:1978vj,Andersson:1990wt,Kozlowski:1994cw,Chaudhuri:2005dt} and multireference coupled cluster methods.\cite{Jeziorski:1981vo,Mahapatra:1998kj,Mahapatra:1999tm,Masik:1998tm,Pittner:1999uk,Hanrath:2005kj,Kong:2009iu,Datta:2011es,Datta:2012hu,Nooijen:2014bx}
The common theme to these approaches is the similarity transformation of the Hamiltonian via a wave operator $\Omega$,  which yields the effective Hamiltonian $H^{\rm eff}$: 
\begin{equation} \label{eq:similarity_transformation}
\Omega: H \rightarrow H^{\rm eff}= \Omega^{-1} H \Omega.
\end{equation}
The similarity transformation is useful to fold in the physical effects of a large number of degrees of freedom into $H^{\rm eff}$, which is defined in a space much smaller than the one spanned by the eigenstates of the original Hamiltonian.
This is achieved by decoupling a set of primary determinants (with associated projector $P$) from the complementary space (with projector $Q = 1 - P$).
For example, coupled cluster theory seeks an exponential parameterization of the wave operator such that $Q H^{\rm eff} P = 0$, while approaches based on a unitary transformation ($\Omega^{-1} = \Omega^\dagger$) simultaneously enforce $QH^{\rm eff}P = 0$ and $PH^{\rm eff}Q = 0$.
This point is illustrated in Fig.~\ref{fig:hamiltonian_decoupling}(A)-(B).

While the importance of the effective Hamiltonian approach cannot be overly emphasized, it is well known that several difficulties are encountered when this formalism is generalized to the case of  multireference electronic states.
Multireference effective Hamiltonian theories are afflicted by two major problems.  The first is the appearance of small energy denominators when a determinant in $Q$ becomes near-degenerate with a determinant in $P$.
This is the well-known intruder-state problem,\cite{Evangelisti:1987fw,Kowalski:2000tx,Kowalski:2000ud}  which is ultimately responsible for convergence problems.
The appearance of small denominators is an intrinsic property of effective Hamiltonian theories, and has been addressed with a variety of techniques.\cite{Malrieu:1985tm,Lepetit:1993te,Roos:1995wf,Assfeld:1995ea,Surjan:1996hg,Finley:2000er,Witek:2002jd,Yanai:2012wp,Stuck:2013bn,Taube:2009jz}
Certain multireference effective Hamiltonian theories also suffer from an imbalance between the number of parameters contained in $\Omega$ and the number of equations that can be derived from the condition $QH^{\rm eff}P = 0$.
For example, in state-specific multireference coupled cluster (MRCC) methods bases on the Jeziorski--Monkhorst ansatz,
\cite{Jeziorski:1981vo,Mahapatra:1998kj,Mahapatra:1998wd,Mahapatra:1999tm,Masik:1998tm,Pittner:1999uk,Hanrath:2005kj,Kong:2009iu} $\Omega$ is overparameterized, and additional sufficiency conditions are imposed to obtain a unique solution. 
In canonical transformation theory,\cite{Yanai:2007ix} the internally-contracted MRCC (ic-MRCC) method,\cite{Mahapatra:1998wd,Evangelista:2011ij,Hanauer:2011io,Evangelista:2012fo,Hanauer:2012hn,Sinha:2013dx} and unitary ic-MRCC approach,\cite{Chen:2012bm} $\Omega$ instead generates linearly-dependent excitations.
While it always possible to formally eliminate the redundant parameters contained in $\Omega$, in practice it is found that convergence problems are unavoidable (see for example Ref.~\onlinecite{Datta:2011es}).
We notice that the redundancy problem can be addressed by replacing the projective conditions with \textit{many-body conditions}, that is by requiring that certain operator components of $H^{\rm eff}$ are zero.\cite{Lindgren:1978tt,Stolarczyk:1985to,Nooijen:1996hp,Nooijen:1997wk,Nooijen:1997kc,Datta:2011es}
To some extent, the many-body conditions maintain a state-specific character, but a perturbative analysis can easily show that the resulting equations are prone to the intruder-state problem.

The similarity renormalization group (SRG) or flow equation method, introduced independently by G{\l}azek and Wilson\cite{Giazek:1993ik,Glazek:1994uu} and Wegner,\cite{Wegner:1994kh} is an alternative approach to transform the Hamiltonian.
The SRG shares the same philosophy of the numerical\cite{Wilson:1975gl} and density-matrix renormalization group,\cite{White:1992tg} but it is based on a continuous unitary version of Eq.~\eqref{eq:similarity_transformation}.
Particularly relevant to this work is the SRG formulated with respect to a Fermi vacuum,\cite{Wegner:2001vs,Kording:2006jr,Wegner:2006cf,Kehrein:2006vz} also called the in-medium SRG.\cite{Tsukiyama:2011eo,Tsukiyama:2012dw,Hergert:2013kf,Hergert:2013hc}
The distinctive feature of the SRG flow  is the decoupling of the Hamiltonian starting from states that have the largest energy separation and progressing to states with smaller energy separation.
Thus, if the many-body basis is sorted according to the energy, a matrix representation of the SRG Hamiltonian is band-diagonal, as shown in Fig.~\ref{fig:hamiltonian_decoupling}(C).

The SRG formalism is particularly attractive because it does not enforce conditions like $QH^{\rm eff}P = 0$, which are at the origin of the intruder state problem.  Instead, states that are degenerate of near-degenerate are not decoupled by the SRG transformation.
Moreover, the SRG is based on a many-body formalism, which addresses the redundancy problem of effective Hamiltonian theories based on projective conditions.

The SRG has been used with great success in the field of nuclear physics to produce soft nucleon-nucleon effective potentials or to obtain the ground-state energy of nuclei.\cite{Bogner:2007hn,Roth:2008hs,Jurgenson:2008gy,Furnstahl:2013eo}
However, applications of the SRG to chemical problems remain largely unexplored, one noticeable exception being an insightful study by White.\cite{White:2002uj}
In this work we explore some potential applications of the SRG to electronic structure theory, both from the formal and computational points of view.
All results presenter in this work are based on a single-reference formalism.
This work has the following goals: 1) to provide an exposition of the SRG approach highlighting its connection to other many-body approaches, 2) to introduce a novel \textit{integral} formulation of the SRG, which we term the driven SRG (DSRG), and 3) to report a numerical comparison of the SRG, the unitary version of DSRG, and the coupled cluster method.

\section{Synopsis of the similarity renormalization group approach} \label{sec:theory_srg}
\subsection{The single-reference SRG formalism}
In this section we will provide a brief introduction to SRG theory and illustrate some of its formal properties.
The focus of this work will be the SRG formulated with respect to a Slater determinant reference $\mref$, built from a one-particle spin orbital basis $\{\phi_p\}$.
The occupied and virtual spin orbitals of $\mref$ will be indicated respectively with the indices $i,j,k,\ldots$ and $a,b,c,\ldots$, while generic spin orbitals will be indicated with the indices $p,q,r,s,\ldots$.
We note that a multireference version of the SRG was recently formulated\cite{Hergert:2013hc} using the extended normal ordering of Mukherjee and Kutzelnigg.\cite{Mukherjee:1997tk,Kutzelnigg:1997ut}
The \textit{bare} Hamiltonian written in second quantized form and normal ordered with respect to $\mref$ is:
\begin{equation} \label{eq:bare_hamiltonian}
H = E_0 + \sum_{pq} \tens{f}{p}{q}\no{\sqop{p}{q}} + \frac{1}{4} \sum_{pqrs} \tens{v}{pq}{rs} \no{\sqop{pq}{rs}},
\end{equation}
where $E_0 = \bra{\mref}H\ket{\mref}$, $\tens{f}{p}{q} = \bra{p}f\ket{q}$, and $\tens{v}{pq}{rs} = \aphystei{pq}{rs}$ are respectively the energy of the reference, the Fock operator integrals, and the antisymmetrized two-electron integrals in physicist notation.  Products of creation ($\hat{a}^\dagger$) and annihilation ($\hat{a}$) operators are written compactly as $\sqop{pq\cdots}{rs\cdots} = \cop{p} \cop{q} \cdots \aop{s} \aop{r}$,\bibnote{
Notice that in the notation used here the top and bottom indices of the matrix elements of the many-body operators are switched with respect to the traditional notation.
For example, the coupled cluster amplitudes are written as $\tens{t}{ab}{ij}$ instead of $\tens{t}{ij}{ab}$.  
However, this way of writing indices reflects the proper tensor character of the many-body operators.
}
and normal ordering with respect to $\mref$ is indicated with curly braces.

In the SRG formalism the Hamiltonian is brought to a diagonal form by a continuous unitary transformation [$U(s)$] parameterized by the flow parameter $s$ defined in the range [0,$\infty)$.  This defines the SRG transformed Hamiltonian [$H(s)$]:
\begin{equation} \label{eq:srg_transformation}
H(s) = U(s) H U^{\dagger}(s).
\end{equation}
In addition, we require that $U(0) = 1$ so that at the beginning of the flow the SRG transformed Hamiltonian is equal to the bare Hamiltonian,  $H(0) = H$.
$H(s)$ is assumed to be expressed in normal-ordered form with respect to $\mref$, and in general it contains three- and higher-body operators:
\begin{equation}
H(s) = E_0(s) + F(s) + V(s) + W(s) + \ldots,
\end{equation}
where $E_0(s)$ is a scalar, $F(s)$ and $V(s)$ are one- and two-body operators,
\begin{align}
F(s) &= \sum_{pq} \tens{f}{p}{q}(s) \no{\sqop{p}{q}},\\
V(s) &= \frac{1}{4} \sum_{pqrs} \tens{v}{pq}{rs}(s) \no{\sqop{pq}{rs}},\\
\end{align}
and $W(s) + \ldots $ stand for three- and higher-body operators.
Notice that the matrix elements of these operators are functions of the flow parameter $s$.

As shown by G{\l}azek and Wilson\cite{Glazek:1994uu} and Wegner,\cite{Wegner:1994kh} the SRG transformation [Eq.~\eqref{eq:srg_transformation}] implies that the Hamiltonian evolves according to the ordinary differential equation (ODE):
\begin{equation} \label{eq:srg_equation}
\frac{d H(s)}{ds} = [\eta(s),H(s)], 
\end{equation}
where the \textit{flow generator}, $\eta(s)$, is an anti-Hermitian operator related to $U(s)$ via the condition:
\begin{equation}
\eta(s) = \frac{d U(s)}{ds}U(s)^{\dagger} = -\eta^\dagger(s).
\end{equation}
The flow variable $s$ has dimensions (energy)$^{-2}$ and may be related to an energy cutoff $\Lambda = s^{-\frac{1}{2}}$, which tends to zero as the SRG equations are propagated.\cite{Kehrein:2006vz}

In the SRG approach, $\eta(s)$ is parameterized in terms of components of $H(s)$, and various choices are possible.\cite{Anderson:2008hm}
For example, the \textit{canonical generator} introduced by Wegner,\cite{Wegner:1994kh} consists in partitioning the Hamiltonian into a diagonal ($H^{\rm d}$) and off-diagonal ($H^{\rm od}$) part, and taking the commutator of $H^{\rm d}(s)$ with $H(s)$:
\begin{equation} \label{eq:canonical_generator}
\eta(s) =  [H^{\rm d}(s),H(s)]  = [H^{\rm d}(s),H^{\rm od}(s)].
\end{equation}
What guarantees the SRG flow will bring $H(s)$ to a diagonal form?
It can be shown\cite{Kehrein:2006vz} that as long as $\eta(s) \neq 0$, the canonical generator reduces the trace of the off-diagonal part of the Hamiltonian, that is:
\begin{equation} \label{eq:srg_trace_condition}
\frac{d}{ds} {\rm Tr}\left[H^{\rm od}(s)^\dagger H^{\rm od}(s) \right] \leq 0.
\end{equation}
Accordingly, as $s\rightarrow \infty$, $E_0(s)$ evolves towards one of the eigenvalues of the original Hamiltonian, while the state $U^{\dagger}(s) \ket{\mref}$ approaches one of its eigenvectors.

The numerical solution of the SRG equations requires integrating Eq.~\eqref{eq:srg_equation} while updating $\eta(s)$.
In practical implementations of the SRG approach it is necessary to truncate $H(s)$ and  $\eta(s)$, otherwise these operators will contain many-body terms of maximum rank equal to the total number of electrons.
Therefore, it is customary to introduce a family of approximate SRG schemes, SRG$(n)$, in which all operators are truncated to rank less than or equal to $n$.  For example, the SRG(2) consists in approximating the SRG equations as:
\begin{align}
\frac{d H(s)}{ds} &= [\eta(s),H(s)]_{1,2}, \label{eq:srg2_hamiltonian}\\
\eta(s) &= [H^{\rm d}(s),H(s)]_{1,2} \label{eq:srg2_eta},
\end{align}
where the commutator subscript (1,2) indicates that only one- and two-body operators normal ordered with respect to the reference $\mref$ are retained.
The SRG(2) approximation is analogous to the singles and doubles truncation scheme used in coupled cluster theory (CCSD), and as will be shown later, these methods have similar accuracy.
We notice that the in-medium SRG(2) approach of Tsukiyama \textit{et al.}\cite{Tsukiyama:2011eo} solves Eqs.~\eqref{eq:srg2_hamiltonian} and \eqref{eq:srg2_eta} but neglects some terms to achieve partial cancellation of three-body contributions to the energy for hard potentials.

\subsection{Perturbative analysis of the SRG equations}
In this section we present a perturbative analysis of the flow equations to illustrate the most important features of the SRG approach.
To this end we partition the initial Hamiltonian ($s = 0$) into a zeroth-order [$H^{(0)}(s=0)$] plus first-order [$H^{(1)}(s=0)$] operator:
\begin{equation}
H(0) = \underbrace{E_0(0) + F(0)}_{H^{(0)}(0)} + \lambda \underbrace{V(0)}_{H^{(1)}(0)},
\end{equation}
where $\lambda$ is a parameter that controls the magnitude of the perturbation, and expand the SRG Hamiltonian into a power series in $\lambda$:
\begin{equation}
H(s) = H^{(0)}(s) + \lambda H^{(1)}(s) + \lambda^2 H^{(2)}(s) + \ldots .
\end{equation}
Following Tsukiyama, Bogner, and Schwenk\cite{Tsukiyama:2011eo} we use the canonical generator and take the off-diagonal part of the Hamiltonian to be:
\begin{equation}
\begin{split} \label{eq:block_wegner}
H^{\rm od}(s) =& \sum_{ia} \tens{f}{i}{a}(s) \no{\sqop{i}{a}} + {\rm H.c.} \\
&+ \frac{1}{4} \sum_{ijab} \tens{v}{ij}{ab}(s) \no{\sqop{ij}{ab}} +{\rm H.c.},
\end{split}
\end{equation}
where H.c. stands for Hermitian conjugate.
This definition of $H^{\rm od}$ leads to a decoupling of the reference determinant from all singly and doubly excited determinants.
Further, we assume to work in a basis of canonical or semicanonical orbitals,\cite{Handy:1989wl} that is, the occupied and virtual orbitals blocks of the Fock matrix are diagonal.  Together with the variation condition, this implies that $\tens{f}{p}{q}(0) = \tens{\delta}{p}{q} \epsilon_p$, where $\epsilon_p = \bra{p}f\ket{p}$.

Perturbative expressions for the SRG using the canonical generator may be easily derived.  Introducing the generalized M{\o}ller--Plessett denominators ($\denominator{ab\cdots}{ij\cdots}$):
\begin{equation} \label{eq:MP_denominators}
\denominator{ab\cdots}{ij\cdots} 
= \epsilon_i + \epsilon_j + \ldots - \epsilon_a - \epsilon_b -\ldots,
\end{equation}
the elements of the first-order generator $\eta^{(1)}(s)$ are given by
\begin{align}
\tens{\eta}{a}{i,(1)}(s) &= - \denominator{a}{i} \,  \tens{f}{a}{i,(1)}(s),\\
\tens{\eta}{ab}{ij,(1)}(s) &= -\denominator{ab}{ij} \,  \tens{v}{ab}{ij,(1)}(s),
\end{align}
and the Hamiltonian flow equations can be written as the set of ODEs for the first-order matrix elements of the Hamiltonian:
\begin{align}
\frac{d}{ds}& E_0^{(1)}(s) = 0,\\
\frac{d}{ds}& \tens{f}{a}{i,(1)}(s) = - \left(\denominator{a}{i} \right)^2  \tens{f}{a}{i,(1)}(s),\\
\frac{d}{ds}& \tens{v}{ab}{ij,(1)}(s) = -\left(\denominator{ab}{ij} \right)^2 \tens{v}{ab}{ij,(1)}(s), \label{eq:model_srg} 
\end{align}
together with a similar set of equations for the matrix elements of $\tens{f}{i}{a,(1)}(s)$ and $\tens{v}{ij}{ab,(1)}(s)$.
Eq.~\eqref{eq:model_srg} may be integrated to yield $E_0^{(1)}(s) = 0$,  $\tens{f}{a}{i,(1)}(s) = 0$, and
\begin{equation} \label{eq:srg_v_pt1}
\tens{v}{ab}{ij,(1)}(s) = \aphystei{ij}{ab} e^{-s \left(\denominator{ab}{ij}\right)^2 }.
\end{equation}
This result illustrates the most important feature of the SRG flow.
In the spirit of a true renormalization theory, the SRG gradually decouples various energy scales in the Hamiltonian.
As $s\rightarrow \infty$, if the energy separation $\epsilon_i + \epsilon_j - \epsilon_a - \epsilon_b$ is not null, $\tens{v}{ab}{ij,(1)}(s)\rightarrow 0$.
For intermediate values of $s$, terms with energy difference $|\denominator{ab}{ij}|$ larger than the energy scale $\Lambda = s^{-\frac{1}{2}}$, are decoupled, while those with smaller energy remain mostly unchanged. 

At the second order of perturbation theory, the flow equation for the energy is:
\begin{equation}
\begin{split}
\frac{d}{ds} E_0^{(2)}(s) = \frac{1}{2} \sum_{ij} \sum_{ab} \denominator{ab}{ij}\left| \tens{v}{ab}{ij,(1)}(s) \right|^2,
\end{split}
\end{equation}
which upon integration over $s$ yields:
\begin{equation} \label{eq:srgpt2-wegner}
E_0^{(2)}(s) = \frac{1}{4} \sum_{ij} \sum_{ab} \frac{|\aphystei{ij}{ab}|^2 \, }{\denominator{ab}{ij} }\left[1 - e^{-2 s \left(\denominator{ab}{ij}\right)^2 } \right].
\end{equation}
A Taylor series expansion of Eq.~\eqref{eq:srgpt2-wegner} with respect to $\denominator{ab}{ij}$ shows that for any finite value of $s$, $E_0^{(2)}(s)$ does not diverge if one of the denominators $\denominator{ab}{ij}$ goes to zero.
This is a beautiful result that highlights another important feature of the SRG: even at the perturbative level the SRG yields a robust second-order energy expression.
Eq.~\eqref{eq:srgpt2-wegner} may also be viewed as a clever way to regularize second-order perturbation theory, which is physically motived and affects mostly only those amplitudes with small denominators.
Thus, the SRG approach is distinct from regularization techniques that increase the gap between occupied and virtual orbitals\cite{Roos:1995wf,Assfeld:1995ea,Surjan:1996hg,Finley:2000er,Witek:2002jd,Yanai:2012wp,Stuck:2013bn} or try to minimize the norm of the amplitudes.\cite{Taube:2009jz}

As a side note, it is important to point out that in numerical applications, the canonical generator usually leads to a stiff set of ODEs, and as suggested by White,\cite{White:2002uj} it is preferable to use an alternative generator parameterized in the following way:
\begin{align}
\label{eq:white_generator_s}
\tens{\eta}{a}{i}(s) &= \frac{ \tens{f}{a}{i}(s) }{\tens{f}{a}{a}(s) - \tens{f}{i}{i}(s) - \tens{v}{ai}{ai}(s)}, \\
\label{eq:white_generator_d}
\tens{\eta}{ab}{ij}(s) &=  \frac{ \tens{v}{ab}{ij}(s) }{\tens{f}{a}{a}(s) + \tens{f}{b}{b}(s) - \tens{f}{i}{i}(s) - \tens{f}{j}{j}(s) + \tens{A}{ab}{ij}(s)},
\end{align}
where $\tens{A}{ab}{ij}(s)=
 \tens{v}{ab}{ab}(s) + \tens{v}{ij}{ij}(s)
- P(ij)\left[\tens{v}{ai}{ai}(s) + \tens{v}{bj}{bj}(s)\right]$.
\bibnote{The matrix elements $\tens{\eta}{i}{a}(s)$ and $\tens{\eta}{ij}{ab}(s)$ are computed from Eqs.~\eqref{eq:white_generator_s} and \eqref{eq:white_generator_d} as $\tens{\eta}{i}{a}(s) = -\tens{\eta}{a}{i}(s)$ and $\tens{\eta}{ij}{ab}(s) = -\tens{\eta}{ab}{ij}(s)$.}
While it is convenient to use the White generator to speedup SRG computations, this generator yields a flow that does not satisfy the traditional definition of renormalization.
This point can be shown by means of a perturbative analysis.   The only contribution to the first-order flow using the White generator yields the ODE $\frac{d}{ds} \tens{v}{ab}{ij,(1)}(s) = -\tens{v}{ab}{ij,(1)}(s)$, which has solution:
\begin{equation} \label{eq:first_order_white}
\tens{v}{ab}{ij,(1)}(s) = \aphystei{ij}{ab} \, e^{-s}.
\end{equation}
Eqs.~\eqref{eq:first_order_white} shows that the White generator leads to a flows in which all off-diagonal elements of the Hamiltonian decay at the same rate. 
Moreover, the second-order energy is given by:
\begin{equation} \label{eq:second_order_white}
E_0^{(2)}(s) = \frac{1}{4} \sum_{ij} \sum_{ab} \frac{|\aphystei{ij}{ab}|^2}{ \denominator{ab}{ij}}\, \left[1 - e^{-2 s} \right].
\end{equation}
This is not a robust expression, since it diverges as M{\o}ller--Plessett perturbation theory when there is an excited determinant for which $\denominator{ab}{ij}\rightarrow 0$.

\section{Theory}
\subsection{Motivations for developing an integral version of the SRG}
As discussed in Section~\ref{sec:theory_srg}, the SRG formalism is an interesting alternative approach to similarity-transform the Hamiltonian: it yields a numerical method that in principle is free from the intruder problem, operates in a many-body formalism, and can be used to selectively eliminate the coupling of the reference with excited configurations down to an energy threshold $\Lambda$.
However, there are some aspects of the SRG that we believe can be improved.
First, it is desirable to have an \textit{integral} formulation of the SRG that does not require to solve a set of differential equations.
In principle the SRG has the same complexity of other many-body approaches like, for example, coupled cluster theory.  However, in practice it is generally more challenging and less numerically robust to solve a set of ODEs than to solve a set of polynomial equations.

Second, it is advantageous to generalize the SRG approach to a formalism similar to coupled cluster theory, in which although it is necessary to truncate the rank of the cluster operator, the ensuing equations do not need to be truncated.
However, it is problematic to setup a SRG based on the coupled cluster-like transformation:
\begin{equation} \label{eq:cc_srg}
H(s) = e^{-T(s)} H e^{T(s)},
\end{equation}
 in which $U(s)$ is replaced by the exponential of the $s$-dependent excitation operator $T(s)$.
Since $\exp[T(s)]$ is not a unitary operator, it appears that a SRG method based on Eq.~\eqref{eq:cc_srg} might not satisfy the trace condition Eq.~\eqref{eq:srg_trace_condition}, which guarantees that the flow of the Hamiltonian decouples the reference from the excited determinants.
Pigg and co-workers\cite{Pigg:2012hu} have described an imaginary-time CC formalism bases on the following set of ODEs:
\begin{equation} \label{eq:it-cc}
\frac{d}{d\tau} \tens{t}{ab\cdots}{ij\cdots}(\tau) = -\bra{\Phi} \sqop{ij\cdots}{ab\cdots}e^{-T(\tau)} H e^{T(\tau)} \ket{\Phi},
\end{equation}
where $\Phi$ is a reference Slater determinant and $T(\tau)$ is an excitation operator function of the imaginary-time variable $\tau$.
However, even this imaginary-time CC formalism is formally problematic.  Indeed, it is possible to show that the first-order imaginary-time CC amplitudes, $\tens{t}{ab}{ij,(1)}(\tau)$, obey the following equation:
\begin{equation}
\tens{t}{ab}{ij,(1)}(\tau) = \frac{\tens{v}{ab}{ij} \left[ 1 - e^{\tau \denominator{ab}{ij}} \right]}{\denominator{ab}{ij}}.
\end{equation}
Contrary to Eq.~\eqref{eq:srg_v_pt1}, the expression for $\tens{t}{ab}{ij,(1)}(\tau)$ can diverge for $\tau\rightarrow \infty$ if there is a double excitation with a positive denominator ($\denominator{ab}{ij} > 0$).

\subsection{The driven similarity renormalization group formalism}
Here we propose to address these two points with a novel SRG method.  The starting point of our \textit{driven similarity renormalization group} (DSRG) approach is the continuous transformation:
\begin{equation} \label{eq:driven_srg}
\bar{H}(s) = e^{-S(s)} H e^{S(s)},
\end{equation}
parameterized by the operator $S(s)$, function of the time-like parameter $s$.
To avoid confusion with the original SRG transformation [Eq.~\eqref{eq:srg_transformation}], we write the DSRG transformed Hamiltonian with a bar above it.
We assume that $S(s)$ can be written as the sum of $k$-body operators [$S_k(s)$], truncated to particle rank $n$:
\begin{equation}
S(s) = \sum_{k=1}^n S_k(s).
\end{equation}
The DSRG ansatz permits us to consider two parameterizations of $S(s)$:
\begin{enumerate}
\item A \textit{coupled cluster or intermediate normalization} DSRG (CC-DSRG), in which $S(s)$ is an excitation operator, $T(s)$.
In this case the $k$-body component of $T(s)$ is a defined as:
\begin{equation} \label{eq:cluster_operator}
T_{k}(s) = \frac{1}{(k!)^{2}}
\sum_{ij\cdots}
\sum_{ab\cdots}
\tens{t}{ab\cdots}{ij\cdots}(s)\,
\sqop{ab\cdots}{ij\cdots},
\end{equation}
where $\sqop{ab\cdots}{ij\cdots}$ is $k$-fold excitation operator.
\item A \textit{unitary} DSRG (for convenience abbreviate simply as DSRG), in which $S(s)$ is an anti-Hermitian operator $A(s)$.  The  $k$-body component of $A(s)$, $A_k(s)$, is expressed in terms of the excitation operators as:
\begin{equation} \label{eq:antiHermitian_operator}
A_{k}(s) = T_{k}(s) -T_{k}^\dagger(s).
\end{equation}
\end{enumerate}

In the DSRG, we postulate that the flow of the similarity Hamiltonian is driven by the \textit{source} operator $R(s)$, according to the following equation:
\begin{equation} \label{dsrg_equation}
[\bar{H}(s)]_{\rm N} =
[e^{-S(s)} H e^{S(s)}]_{\rm N} = R(s) \;\;\;\;\;s \in [0,\infty),
\end{equation}
where the subscript N indicates the \textit{non-diagonal}\cite{Kutzelnigg:2010uv} diagrams of $\bar{H}(s)$.
A distinction must be made depending on the parameterization of $S(s)$:
\begin{enumerate}
\item In the coupled cluster parameterization of the DSRG, the non-diagonal part of $\bar{H}(s)$ is defined as its excitation component.  In this case $R$ is a pure excitation operator, $R = R_{\rm ex}$, and $R_{\rm ex}$ has the same structure as $T$, but it is parameterized by the amplitudes $\tens{r}{ab\cdots}{ij\cdots}(s)$.   
\item In the unitary DSRG, the non-diagonal part of an operator is the sum of its pure excitation and deexcitation components.  In this case $R$ is Hermitian, and it may be expressed as the sum of excitation and deexcitation operators, $R = R_{\rm ex} + R_{\rm ex}^\dagger$.
\end{enumerate}

The DSRG flow equation [Eq.~\eqref{eq:driven_srg}] must be augmented with appropriate  boundary conditions for the operator $R(s)$.
For $s=0$ we demand that the non-diagonal component of the Hamiltonian is identical to the bare Hamiltonian:
\begin{equation} \label{eq:dsrg_boundary_zero}
[\bar{H}(0)]_{\rm N} = H_{\rm N},
\end{equation}
a condition that can be trivially satisfied if $S(0) = 0$.  Furthermore, Eq.~\eqref{eq:dsrg_boundary_zero} implies that
\begin{equation} \label{eq:r_zero_boundary_condition}
R(0) = R_1(0) + R_2(0) = H_{\rm N}.
\end{equation}
For $s\rightarrow \infty$, we require that the DSRG flows decouples excited configurations from the reference, driving the non-diagonal part of $\bar{H}$ to zero, $[\bar{H}(\infty)]_{\rm N} = 0$.  Thus, the appropriate boundary condition for the matrix elements of $R(s)$ is:
\begin{equation} \label{eq:r_infty_boundary_condition}
\lim_{s\rightarrow \infty}\tens{r}{ab\cdots}{ij\cdots}(s) =
0.
\end{equation}
Notice that other boundary conditions for the limit $s\rightarrow \infty$ may be more advantageous.  For example, one may require the DSRG transformation to avoid decoupling excited configurations for which $\denominator{ab\cdots}{ij\cdots} = 0$.  This boundary condition will not guarantee that the DSRG transformation will diagonalize the Hamiltonian, but it will make the DSRG equations numerically stable in the presence of degenerate excited determinants.

A hierarchy of truncated DSRG methods that converges to the FCI method can be formulated by limiting the particle rank of the $S$ operator.  These will be abbreviated with DSRG($n$), where $n$ indicates the highest particle rank of $S$.
In this work we will consider the unitary DSRG singles and doubles [DSRG(2)], defined by $S(s) = T_1(s) + T_2(s) - T_1^\dagger(s) - T_2^\dagger(s)$.
In addition, the unitary DSRG also requires approximating the transformed Hamiltonian $\bar{H}$.\bibnote{This problem is analogous to the one encountered in unitary coupled cluster theory, where it is necessary to truncate the Baker--Campbell--Hausdorff expansion for $\bar{H}$.   However, in the coupled cluster version of DSRG, the similarity-transformed Hamiltonian truncates after four commutators, and thus, it is not necessary to approximate the corresponding amplitude equations.}  In this work we use the approximate Baker--Campbell--Hausdorff (BCH) formula suggested by Yanai and Chan\cite{Yanai:2006gi} to compute $\bar{H}$ truncated to one- and two-body operators ($\bar{H}_{1,2}$):
\begin{equation} \label{eq:CT_BCH}
\bar{H}_{1,2} =  H + \sum_{k=1}^\infty 
\frac{1}{k!}\underbrace{[\cdots[[H,A]_{1,2},A]_{1,2},\cdots]_{1,2}}_{k\text{-fold commutator}}.
\end{equation}
In the unitary DSRG(2), the flow equation [Eq.~\eqref{dsrg_equation}] reads as a set of four equations:
\begin{equation} \label{eq:dsrgsd}
\begin{cases}
\tens{[\bar{H}(s)]}{a}{i} =\tens{r}{a}{i}(s) & \tens{[\bar{H}(s)]}{i}{a} =\tens{r}{i}{a}(s) \\
\tens{[\bar{H}(s)]}{ab}{ij} =\tens{r}{ab}{ij}(s) & \tens{[\bar{H}(s)]}{ij}{ab} =\tens{r}{ij}{ab}(s),
\end{cases}
\end{equation}
where some of the $r$ amplitudes are related via Hermitian conjugation: $\tens{r}{i}{a}(s) = [\tens{r}{a}{i}(s)]^*$ and $\tens{r}{ij}{ab}(s) = [\tens{r}{ab}{ij}(s)]^*$.
When it comes to specify a form for the source operator, the boundary conditions Eqs.~\eqref{eq:dsrg_boundary_zero} and \eqref{eq:r_infty_boundary_condition} leave a considerable amount of freedom.
In this work we use the results of the perturbative analysis of the SRG equations [Eq.~\eqref{eq:srg_v_pt1}] to suggests the following parameterization of $R(s)$:
\bibnote{
First-order perturbation theory does not provide insight into the form of the triples and higher amplitude equations, and several options are possible.
One possibility is a generalization of the singles and doubles equations for $R(s)$: 
\begin{equation}
\tens{r}{ab\cdots}{ij\cdots}(s) = \, \left[\tens{\bar{H}}{ab\cdots}{ij\cdots}(s) + \tens{t}{ab\cdots}{ij\cdots} \denominator{ab\cdots}{ij\cdots}\right] \, e^{-s (\denominator{ab\cdots}{ij\cdots})^2},
\end{equation}
but the simpler condition:
\begin{equation}
\tens{r}{ab\cdots}{ij\cdots}(s) = 0,
\end{equation}
also appears to be viable.
}
\begin{align} \label{eq:r_canonical_one}
\tens{r}{a}{i}(s) & = \, \left[\tens{\bar{H}}{a}{i}(s) + \tens{t}{a}{i} \denominator{a}{i}  \right]  \, e^{-s (\denominator{a}{i})^2},\\ \label{eq:r_canonical_two}
\tens{r}{ab}{ij}(s) &= \, \left[\tens{\bar{H}}{ab}{ij}(s) + \tens{t}{ab}{ij} \denominator{ab}{ij}\right] \, e^{-s (\denominator{ab}{ij})^2}.
\end{align}

To highlight the connection between our choice of the source operator $R(s)$ and the similarity-renormalization group we expand the source operator up to first order in $\lambda$, and observe that the matrix elements of $\bar{H}$, in the case of a Hartree--Fock reference, evolve like those of the SRG:
\begin{align} \label{eq:r_canonical_pt1}
\tens{\bar{H}}{a}{i,(1)}(s) & = 0,\\
\tens{\bar{H}}{ab}{ij,(1)}(s) &= \aphystei{ij}{ab} \, e^{-s (\denominator{ab}{ij})^2}.
\end{align}
Similarly, the first-order DSRG single- and double-excitation amplitudes are given by: 
\begin{align}
\tens{t}{a}{i,(1)} =& \, 0,\\
\tens{t}{ab}{ij,(1)} =& \, \frac{\tens{v}{ab}{ij} \left[ 1 - e^{-s (\denominator{ab}{ij})^2} \right]}{\denominator{ab}{ij}}.
\end{align}
The doubles amplitudes $\tens{t}{ab}{ij,(1)}$ have a renormalized denominator that for any finite and non zero $s$ guarantees that $|\tens{t}{ab}{ij,(1)}| < \infty$ even in the limit $\denominator{ab}{ij}\rightarrow 0$.
It is important to emphasize that alternative choices of the $R$ operator are possible.  For example, in our initial study we considered another form of $R(s)$ compatible with the perturbative analysis: $\tens{r}{a}{i}(s) = \, \tens{f}{a}{i}\exp[-s (\denominator{a}{i})^2]$  and $\tens{r}{ab}{ij}(s) = \, \aphystei{ab}{ij}\exp[-s (\denominator{ab}{ij})^2]$.  This is a more obvious choice, but Eqs.~\eqref{eq:r_canonical_one} and \eqref{eq:r_canonical_two} lead to equations that are numerically more robust.

\subsection{Formal comparison of the SRG, DSRG, and other approaches}
\label{sec:formal_comparison}
\begin{figure}[t]
\begin{center}
\includegraphics[width=3.5in]{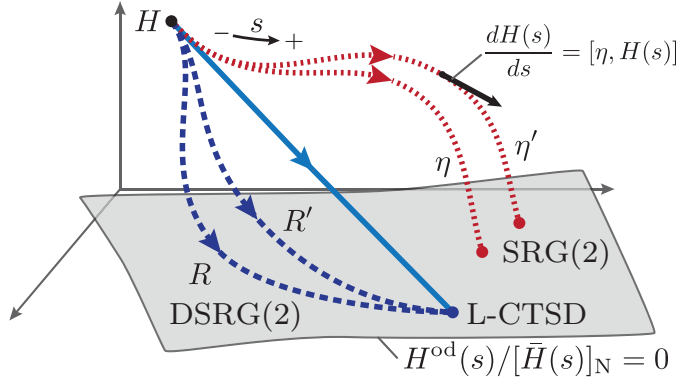}
\caption{Illustration of the relationship between truncated versions of the similarity renormalization group approach, the driven SRG, and linearized canonical transformation theory.  Starting from the bare Hamiltonian, all methods bring the non-diagonal component of the transformed Hamiltonian ($H^{\rm od}(s)$ or $[\bar{H}(s)]_{\rm N}$) to zero.  The SRG(2) flow follows the derivative induced by the generator $\eta$, and in the limit $s \rightarrow \infty$ different generators lead to different transformed Hamiltonians.  The DSRG(2) flow is also dependent on the source operator $R$, but as long as two source operators satisfy the same boundary conditions the final transformed Hamiltonian is identical to the L-CTSD one.}
\label{fig:comparison}
\end{center}
\end{figure}
In this section we will explore formal connections between the unitary version of SRG, the DSRG, and other unitary formalisms like canonical transformation theory\cite{White:2002uj,Yanai:2006gi,Yanai:2007ix} and the anti-Hermitian contracted Schr\"{o}dinger equation (ACSE).\cite{Mazziotti:2006iw,Mazziotti:2007gs,Mazziotti:2012dx}
Let us begin with a comparison of the DSRG and the SRG methods.
As illustrated in Fig.~\ref{fig:comparison} both approaches generate a flow of the original Hamiltonian as a function of the parameter $s$.  In the SRG, the flow of the Hamiltonian follows the path induced by the generator $\eta$.
Generally speaking, in absence of truncation and degeneracies, the SRG method is exact, and in the limit $s \rightarrow \infty$ the transformed Hamiltonian is independent of the particular generator used.
However, as pointed out in Fig.~\ref{fig:comparison}, when the SRG equations are approximated, different generators $\eta$ and $\eta'$  will induce different paths that are not guaranteed to converge to the same final Hamiltonian.

The distinctive feature of the DSRG, is that the flow of the Hamiltonian is driven by the operator $R$, which is parameterized and \textit{memoryless}, that is, $R(s)$ does not depend on $R(s')$ at earlier times ($s' < s$).
Thus, for a fixed value of $s$, the DSRG method consists of a set of polynomial equations, for which efficient solvers and convergence accelerators have been developed.\cite{Pulay:1980jn,Scuseria:1986tn}
As in the case of the SRG based on the canonical generator, the DSRG is designed to progressively renormalize the coupling between the reference and the excited configurations that can interact with it.
Consequently, the DSRG limits the decoupling of excited determinants with small energy denominators and naturally restraints the magnitude of the amplitudes entering the $S$ operator.
Fig.~\ref{fig:comparison} also illustrates the path-independence of the DSRG transformed Hamiltonian.  In the limit $s \rightarrow \infty$, two source operators $R(s)$ and $R'(s)$ that satisfy the boundary conditions Eqs.~\eqref{eq:dsrg_boundary_zero} and \eqref{eq:r_infty_boundary_condition}  yield the same DSRG Hamiltonian.

The $s \rightarrow \infty$ limit of the intermediate normalization and unitary versions of the DSRG is of special interests because these formalisms then become respectively equivalent to coupled cluster theory and unitary coupled cluster theory/canonical transformation theory.\cite{Kutzelnigg:1982wv,Bartlett:1989th,White:2002uj,Yanai:2006gi,Yanai:2007ix}
When $s \rightarrow \infty$ the unitary DSRG(2) equation reads
\begin{equation}
[\bar{H}_{1,2}]_{\rm N} = 0,
\end{equation}
which in the single-reference case is equivalent to the linearized canonical transformation (L-CT) condition:\cite{Yanai:2006gi}
\begin{equation}
\bra{\Phi}[\bar{H}^\text{L-CT}_{1,2},\sqop{ab\cdots}{ij\cdots} - \sqop{ij\cdots}{ab\cdots}]\ket{\Phi}=0,
\end{equation}
where $\bar{H}^\text{L-CT}_{1,2}$ is the CT Hamiltonian, $\bar{H}^{\rm CT} =  e^{-A^{\rm CT}} H  e^{A^{\rm CT}}$, approximated via Eq.~\eqref{eq:CT_BCH} and the anti-Hermitian operator $A^{\rm CT}$ is defined analogously to Eq.~\eqref{eq:antiHermitian_operator}.

The comparison of the SRG and canonical transformation theory is also instructive.  Formally, the \textit{exact} SRG unitary transformation $U^\dagger(s)$ can be expressed as the exponential of an anti-Hermitian operator $A(s)$, $U^\dagger(s) = \exp[{A(s)}]$, which is reminiscent of the DSRG and CT transformed Hamiltonian.
This observation does not provide much insight into the relationship of the \textit{approximate} SRG(2) and L-CTSD methods, which are based on different commutator approximations.
We numerically confirmed that the SRG(2) with the canonical or White generator and the unitary DSRG(2) are not equivalent.
However, it is possible to go one more step further and show that the commutator approximation in the SRG(2) may be viewed as a continuous analogous of the one used in the DSRG and L-CT methods.  First we notice that the differential equation for the SRG(2) flow [Eq.~\eqref{eq:srg2_hamiltonian}] is equivalent to the following integral equation with an approximate commutator:
\begin{equation} \label{eq:srg2_integral}
H(s) = H(0) + \int_0^s ds_1 [H(s_1),\delta A^{\rm SRG}(s_1)]_{1,2},
\end{equation}
where we have introduced the operator $\delta A^{\rm SRG}(s) = -\eta(s)$.
The iterative solution of Eq.~\eqref{eq:srg2_integral} yields a Dyson series in which commutators are truncated to one- and two-body terms.  This series may be written in $s$-ordered form as:
\begin{widetext}
\begin{equation} \label{eq:SRG_Dyson}
H(s) = H(0) +  \sum_{k = 1}^{\infty}
 \frac{1}{k!}
\int_0^s ds_k \cdots \int_0^s ds_1\,
\underbrace{ \mathcal{T}_s \Big\{[\cdots [[H(0),\delta A^{\rm SRG}(s_1)]_{1,2},\delta A^{\rm SRG}(s_2)]_{1,2}, \cdots ]_{1,2}}_{k\text{-fold commutator}}\Big\},
\end{equation}
\end{widetext}
where $\mathcal{T}_s$ is the $s$-ordering operator, which rearranges the $\delta A^{\rm SRG}$ terms in the integral by a   permutation ($\pi$) of the indices of $s$, such that $s_{\pi_1} \leq s_{\pi_2} \leq \cdots\leq s_{\pi_n}$.
Thus, the SRG(2) commutator approximation [Eq.~\eqref{eq:SRG_Dyson}] is essentially a continuous version of the truncated BCH series [Eq.~\eqref{eq:CT_BCH}] used in CT theory.
In the former approach, $\delta A^{\rm SRG}(s) = -\eta(s)$ plays the continuous counterpart of $A^{\rm CT}$ in the discrete CT transformation.

Some computational aspects of the SRG and the DSRG are also similar to the idea of parameterizing the exact wave function in terms of unitary rotation of a reference $N$-representable wave function\cite{Kutzelnigg:2003ft,Kutzelnigg:2004ep} as done in the anti-Hermitian contracted Schr\"{o}dinger equation (ACSE) approach of Mazziotti.\cite{Mazziotti:2006iw,Mazziotti:2007gs,Mazziotti:2012dx}
The ACSE method considers a sequence of infinitesimal two-body unitary transformations of an initial wave function $\ket{\Psi(\lambda)}$:
\begin{equation}
\ket{\Psi(\lambda + \delta\lambda} = e^{\delta\lambda S(\lambda)} \ket{\Psi(\lambda)},
\end{equation}
where $S(\lambda)$ is a general two-body anti-Hermitian operator: $S(\lambda) = \sum_{pqrs} \tens{S}{pq}{rs}(\lambda) \sqop{pq}{rs}$.
The feature that distinguishes the ACSE method from the SRG and DSRG approaches is the propagation of the two-particle reduced density matrix, in addition to to the propagation of the energy and $S(\lambda)$.
A stronger connection between the ACSE and the SRG may be established if one considers a more general SRG approach in which the reference state used to define normal ordering and the definition of normal ordering are also considered dependent on $s$.\cite{Kording:2006jr}

It is also worth pointing out that the methods considered here are not related to the family of renormalized coupled cluster methods proposed by Kowalski and Piecuch.\cite{Kowalski:2000ur,Kowalski:2000wj,Piecuch:2002vx,Piecuch:2005tl}
These renormalized coupled cluster methods are based on an exact correction of the coupled cluster energy, expressed in terms of the generalized moments of the coupled cluster equations and amplitudes for the excitation operators or left eigenstates.

\section{Implementation and computational details}
The SRG(2) and the unitary version of DSRG(2) are implemented as a plugin to the quantum chemistry package \textsc{Psi4}.\cite{Turney:2011gr}
The SRG(2) equations [Eqs.~\eqref{eq:srg2_hamiltonian} and \eqref{eq:srg2_eta}] are evaluated using the commutator \textit{keeping all the one- and two-body terms}.
The appendix reports the matrix elements of the commutator $C = [A,B]_{1,2}$, where $A$, $B$, and $C$ are three generic operators truncated to one- and two-body terms and expressed in a normal-ordered form with respect to the reference.
The SRG flow equation is integrated using the eighth-order Runge--Kutta--Fehlberg method with seventh-order error estimation\cite{Fehlberg:1968tz} as implemented in the boost C++ library.\cite{Boost:uw}
Our implementation can perform SRG computations using the White generator or the block-diagonal form\cite{Tsukiyama:2011eo} of the canonical generator as defined in Eq.~\eqref{eq:block_wegner}.
The computational cost required to evaluate the commutator $[H(s),\eta(s)]_{1,2}$, is dominated by a contribution proportional to $O^2 V^2 N^2$, where $O$ and $V$ are respectively the number of occupied and virtual spin orbitals, and $N = O + V$.
Thus the computational scaling of the SRG(2) equations is marginally higher than that of the CCSD method ($O^2 V^4$).

In the case of the DSRG(2) method, the matrix elements of the similarity-transformed Hamiltonian are computed using the approximate BCH expansion given in Eq.~\eqref{eq:CT_BCH} by means of the recursive expression
\begin{equation}
\left\{
\begin{split}
C^{(k+1)} &= \frac{1}{k+1} [C^{(k)},A]_{1,2},\\
\bar{H}^{(k+1)} &= \bar{H}^{(k)} + C^{(k+1)},\\
\end{split}
\right.
\end{equation}
starting with $C^{(0)} = \bar{H}^{(0)} = H$.
The computational scaling of the DSRG(2) approach is identical to that of the SRG(2) since they are based on the same commutator approximation.

All results are obtained using the correlation-consistent valence triple zeta basis set (cc-pVTZ)\cite{Dunning:1989uk} and restricted-Hartree--Fock (RHF) orbitals.  Moreover, all computations of the correlation energy are performed by dropping the 1s-like orbital of the atoms Li trough F.
Equilibrium geometries, harmonic vibrational frequencies, and anharmonic force constants of the diatomic molecules are computed from an equally-spaced five-point polynomial fit of the potential energy curve centered around the CCSD(T) equilibrium bond length with points separated by 0.01 \AA{}.  The harmonic vibrational frequencies and anharmonic force constants are computed using the equations reported in Ref.~\onlinecite{Crawford:1996tb}.

\section{Results}
\subsection{Evolution of the SRG(2) and DSRG(2) energy}
\begin{figure}[ht]
\begin{center}
\includegraphics[width=3.5in]{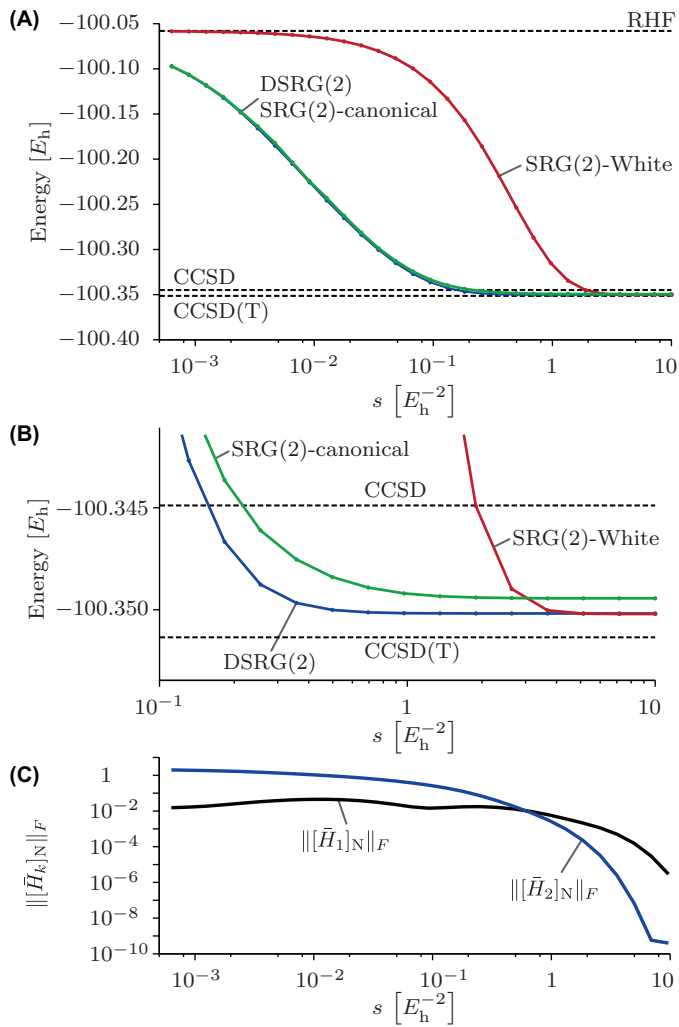}
\caption{Evolution of the SRG(2) and DSRG(2) energy as a function of $s$ (plotted on a logarithmic scale) for the ground electronic state of hydrogen fluoride. 
The figure also reports the energy computed at the restricted-Hartree--Fock, CCSD, and CCSD(T) levels of theory.  Energy evolution in the range $s \in [10^{-3},10]$ (A) and $s \in [0.1,10]$  (B).
(C) Evolution of the Frobenius norm of the $k$-body non-diagonal components of the DSRG transformed Hamiltonian ($\|[\bar{H}_k]_{\rm N}\|_F$).
All results were computed using the cc-pVTZ basis set and RHF orbitals, freezing the fluorine 1s-like molecular orbital in the computations of the correlation energy.}
\label{fig:evolution_hf_re}
\end{center}
\end{figure}

In this section we compare the evolution of the SRG(2) and DSRG(2) energy as a function of $s$.
For the SRG(2) we employ both the canonical and White generators.
Fig.~\ref{fig:evolution_hf_re}(A) shows the evolution of $E_0(s)$ computed for hydrogen fluoride at the experimental equilibrium distance ($r_{\rm e}$ = 0.9168 \AA{}, from Ref.~\onlinecite{Herzberg:i0hQ0hVT}).
The SRG(2) flow obtained using the canonical and White generators is significantly different.  The most clear disparity is the rate of convergence of $E_0(s)$ for these two variants of the SRG(2). 
This result can be understood using the perturbative analysis presented in the previous section.  The White generator yields a flow of the form $E_0(s) \approx E_0(0) + [E_0(\infty) - E_0(0)] (1-e^{-2s})$, while the canonical generator yields a flow in which each contribution decays with a rate proportional to the square of the corresponding denominator, that is $1-e^{-2s (\denominator{ab}{ij})^2}$ [see Eq.~\eqref{eq:srgpt2-wegner}].  Since most denominators will be larger than 1 \Eh, one expects the canonical generator to converge faster to the asymptotic energy value.
Interestingly, for $s \rightarrow \infty$, the SRG energy for both variants converges to a value that is close to the CCSD(T) energy.
This phenomenon was also observed in nuclear structure computations,\cite{Hergert:2013kf} but it is unclear if it is an indication that the SRG(2) truncation scheme is superior to CCSD.
The DSRG(2) flow is almost indistinguishable from the one obtained from the canonical generator/SRG(2), but on a finer scale it is possible to notice a difference between the two approaches.
Also, in Fig.~\ref{fig:evolution_hf_re}(B) the DSRG(2) and White generator/SRG(2) curves appear to tend to the same limit, but the two methods yield energies that differ by about $7 \times 10^{-4}$ \Eh.

An important difference between these methods is the computational time required to evaluate $E(s)$.
For the example considered here, computing $E(s = 10 \,E_{\rm h}^{-2})$ requires seven iterations of the  DSRG(2) equations using the direct inversion of the iterative subspace (DIIS) to speed up convergence.\cite{Pulay:1980jn,Scuseria:1986tn}  The SRG(2) with the White generator has a computational cost that is comparable to the DSRG(2), requiring 41 steps in the integration of the corresponding set of ODEs.
At the same time we confirm that the SRG(2) with canonical generator leads to an inefficient numerical scheme that requires several thousands steps to achieve convergence of the ODEs.
For this reason, all SRG(2) results presented in the following sections will only use the White generator.

Last, we consider the flow of the off-diagonal elements of the Hamiltonian in the DSRG method.
Fig.~\ref{fig:evolution_hf_re}(C) shows the Frobenius norm of the $k$-body non-diagonal components of $\bar{H}$, $\|[\bar{H}_k]_{\rm N}\|_F$, defined as:
\begin{equation}
\|[\bar{H}_k]_{\rm N}\|_F = \sqrt{ \frac{1}{(k!)^2} \sum_{ij\cdots} \sum_{ab\cdots} \left( |\tens{\bar{H}}{ab\cdots}{ij\cdots}(s)|^2 + |\tens{\bar{H}}{ij\cdots}{ab\cdots}(s)|^2 \right)}.
\end{equation}
This plots shows that the DSRG behaves like a proper renormalization theory.  As $s$ is increased, the operator $S(s)$ gradually decouples the reference from the singly and doubly excited determinants.
This suggests that one might perform a DSRG transformation with a finite value of $s$ that is sufficiently large to recover dynamical correlation effects and produce a renormalized Hamiltonian ($\bar{H}$).  By changing $s$, it is possible to tune the norm of the renormalized Hamiltonian, that is, the strength of the residual interactions not accounted by the DSRG transformation.

\subsection{Diatomic molecules}
In this section we evaluate the accuracy of the SRG(2) (White generator) and DSRG(2) methods by computing the equilibrium properties of a series of diatomic molecules for fixed values of the parameter $s$.
Table~\ref{tab:diatomics} reports the equilibrium bond length ($r_{\rm e}$), the harmonic vibrational frequency ($\omega_{\rm e}$), and the correlation energy computed at the CCSD(T) geometry ($\epsilon$), expressed as differences with respect to the CCSD(T) values, for the ground electronic state of H$_2$, LiH, BeO, BH, HF, LiF, BF, C$_2$, N$_2$, O$_2$, F$_2$, and CN$^{-}$.
Table~\ref{tab:diatomics_errors} reports error metrics for the results in Table~\ref{tab:diatomics}, including results for the anharmonic force constant ($\omega_{\rm e} x_{\rm e}$).

\begingroup
\squeezetable
\begin{table*}[tbp]
\centering
\caption{\scriptsize 
Equilibrium properties for the ground electronic state of various diatomic molecules.  Equilibrium bond length ($r_{\rm e}$) in \AA{}, and equilibrium harmonic frequency ($\omega_{\rm e}$) in cm$^{-1}$, and energy error with respect to CCSD(T) at the CCSD(T) equilibrium geometry ($\epsilon$) in m\Eh.  All results were computed using the cc-pVTZ basis set and restricted Hartree--Fock orbitals, freezing the Li--F 1s-like molecular orbitals in the computation of the correlation energy.}
\label{tab:diatomics}
\begin{tabular*}{6.5in}{@{\extracolsep{\fill}}lcr@{}r@{}r@{}r@{}r@{}r@{}r@{}r@{}r@{}r@{}}
\hline\hline
Molecule&Property & SRG(2) & SRG(2*) & \dsrg & \dsrg & \dsrg  & \dsrgph & \dsrgph & \dsrgph & CCSD & CCSD(T) \\
& & \multicolumn{1}{c}{$s$ = 5} & \multicolumn{1}{c}{$s$ = 5}  & \multicolumn{1}{c}{$s$ = 1}  & \multicolumn{1}{c}{$s$ = 10} & \multicolumn{1}{c}{$s$ = $\infty$} & \multicolumn{1}{c}{$s$ = 1} & \multicolumn{1}{c}{$s$ = 10} & \multicolumn{1}{c}{$s$ = $\infty$}  & \\
\hline
H$_2$ & $r_{\rm e}               $ &     0.0018 &    -0.0001 &     0.0013 &     0.0019 &     0.0019 &    -0.0004 &     0.0000 &     0.0000 &     0.0000 &     0.7426 \\
& $\omega_{\rm e}          $ &      -40.9 &        2.4 &      -24.1 &      -43.7 &      -43.7 &       15.2 &       -0.3 &       -0.3 &        0.0 &     4409.4 \\
& $\epsilon                $ &       -1.0 &        0.0 &       -0.9 &       -1.0 &       -1.0 &       -0.1 &       -0.0 &       -0.0 &        0.0 &  -1.172337 \\[6pt]
LiH & $r_{\rm e}               $ &     0.0030 &    -0.0004 &    -0.0084 &     0.0032 &     0.0035 &    -0.0091 &    -0.0002 &    -0.0001 &     0.0000 &     1.6081 \\
& $\omega_{\rm e}          $ &      -13.9 &        2.8 &       36.5 &      -15.5 &      -17.5 &       40.0 &        1.3 &        0.6 &        0.0 &     1394.9 \\
& $\epsilon                $ &       -1.3 &       -0.1 &        0.9 &       -1.4 &       -1.5 &        1.9 &        0.0 &        0.0 &       -0.0 &  -8.022320 \\[6pt]
BeO & $r_{\rm e}               $ &    -0.0065 &    -0.0163 &    -0.0192 &    -0.0068 &    -0.0059 &    -0.0239 &    -0.0164 &    -0.0161 &    -0.0158 &     1.3443 \\
& $\omega_{\rm e}          $ &       41.5 &       96.4 &      123.3 &       45.3 &       39.6 &      145.8 &       98.7 &       96.8 &      101.0 &     1458.8 \\
& $\epsilon                $ &        4.7 &       16.9 &       10.7 &        4.5 &        4.2 &       17.7 &       16.7 &       16.6 &       16.0 & -89.748155 \\[6pt]
BH & $r_{\rm e}               $ &     0.0012 &    -0.0009 &     0.0044 &     0.0002 &     0.0002 &     0.0004 &    -0.0011 &    -0.0011 &    -0.0011 &     1.2354 \\
 & $\omega_{\rm e}          $ &      -14.8 &       11.7 &      -17.7 &      -12.4 &      -12.4 &       15.4 &       10.2 &       10.2 &        9.7 &     2350.8 \\
 & $\epsilon                $ &       -5.4 &        2.4 &        4.8 &       -6.1 &       -6.2 &        8.8 &        2.0 &        1.2 &        2.0 & -25.230615 \\[6pt]
HF & $r_{\rm e}               $ &     0.0005 &    -0.0022 &     0.0006 &     0.0007 &     0.0007 &    -0.0022 &    -0.0021 &    -0.0021 &    -0.0021 &     0.9172 \\
 & $\omega_{\rm e}          $ &      -17.3 &       39.9 &      -17.6 &      -21.7 &      -21.7 &       38.8 &       35.9 &       35.9 &       34.7 &     4177.4 \\
 & $\epsilon                $ &       -9.7 &        2.4 &        5.4 &      -17.9 &      -18.4 &       14.9 &       12.0 &       12.0 &       11.6 & -100.196873 \\[6pt]
LiF & $r_{\rm e}               $ &    -0.0006 &    -0.0028 &    -0.0019 &    -0.0006 &    -0.0006 &    -0.0037 &    -0.0028 &    -0.0027 &    -0.0026 &     1.5836 \\
 & $\omega_{\rm e}          $ &        1.5 &        5.9 &        4.6 &        1.5 &        1.5 &        8.0 &        5.8 &        5.7 &        5.4 &      908.8 \\
 & $\epsilon                $ &        2.0 &        7.9 &        2.7 &        2.0 &        2.0 &        8.4 &        7.8 &        7.8 &        7.7 & -107.263654 \\[6pt]
BF & $r_{\rm e}               $ &     0.0013 &    -0.0035 &    -0.0043 &     0.0014 &     0.0014 &    -0.0069 &    -0.0035 &    -0.0035 &    -0.0034 &     1.2715 \\
 & $\omega_{\rm e}          $ &       -6.7 &       15.4 &       16.8 &       -6.2 &       -6.1 &       29.6 &       16.0 &       16.0 &       15.3 &     1402.7 \\
 & $\epsilon                $ &        0.6 &        4.9 &        7.1 &        0.4 &        0.4 &       15.8 &       11.2 &        7.7 &       11.0 & -124.501909 \\[6pt]
C$_2$ & $r_{\rm e}               $ &        &        &     0.0031 &        &        &    -0.0006 &    -0.0041 &    -0.0041 &    -0.0038 &     1.2507 \\
 & $\omega_{\rm e}          $ &     &     &       54.9 &     &     &       61.8 &       43.7 &       42.6 &       35.7 &     1845.6 \\
 & $\epsilon                $ &         &    &       35.2 &    &    &       60.8 &       36.9 &       20.9 &       33.6 & -75.783164 \\[6pt]
N$_2$ & $r_{\rm e}               $ &     0.0015 &    -0.0071 &    -0.0010 &     0.0020 &     0.0020 &    -0.0087 &    -0.0073 &    -0.0073 &    -0.0070 &     1.1038 \\
 & $\omega_{\rm e}          $ &      -28.6 &       80.0 &       24.5 &      -41.8 &      -41.8 &      109.4 &       80.9 &       80.9 &       77.9 &     2346.0 \\
 & $\epsilon                $ &        2.0 &       18.8 &        2.9 &        1.8 &        1.8 &       19.8 &       19.0 &       19.0 &       18.7 & -109.373937 \\[6pt]
O$_2$ (X $^3\Sigma_{\rm g}^{-}$) & $r_{\rm e}               $ &     0.0023 &    -0.0116 &    -0.0021 &     0.0015 &     0.0015 &    -0.0140 &    -0.0125 &    -0.0125 &    -0.0124 &     1.2121 \\
 & $\omega_{\rm e}          $ &      -34.7 &       81.9 &       17.3 &      -26.8 &      -26.8 &      108.2 &       90.6 &       90.6 &       89.5 &     1585.4 \\
 & $\epsilon                $ &        2.9 &       19.3 &        4.4 &        3.3 &        3.3 &       22.9 &       20.0 &       20.0 &       19.9 & -150.128774 \\[6pt]
F$_2$ & $r_{\rm e}               $ &       $\infty$(?) &     $\infty$(?) &     0.0398 &       $\infty$(?) &      $\infty$(?) &    -0.0268 &    -0.0230 &    -0.0230 &    -0.0211 &     1.4158 \\
 & $\omega_{\rm e}     $ &                        &                       &     -156.3 &                           &                    &      139.7 &      102.1 &      102.1 &       92.5 &      919.9 \\
 & $\epsilon                $ &       -2.9 &       18.5 &       -2.6 &       -3.8 &       -3.8 &       19.3 &       18.8 &       18.8 &       18.1 & -199.296112 \\[6pt]
CN$^{-}$ & $r_{\rm e}               $ &     0.0001 &    -0.0072 &    -0.0050 &     0.0003 &     0.0003 &    -0.0102 &    -0.0074 &    -0.0074 &    -0.0071 &     1.1833 \\
 & $\omega_{\rm e}          $ &       -5.4 &       63.9 &       62.2 &       -9.8 &       -9.8 &      102.3 &       65.6 &       65.6 &       63.6 &     2068.5 \\
 & $\epsilon                $ &        3.5 &       18.7 &        6.1 &        3.3 &        3.3 &       20.6 &       18.7 &       18.7 &       18.4 & -92.694076 \\
\hline\hline
\end{tabular*}
\end{table*}
\endgroup

\begingroup
\squeezetable
\begin{table*}[tbp]
\centering
\caption{
Error analysis of the equilibrium properties for the ground electronic state of various diatomic molecules computed using the cc-pVTZ basis set.  Equilibrium bond length ($r_{\rm e}$) in \AA{}, harmonic frequency ($\omega_{\rm e}$) and anharmonic force constant ($\omega_{\rm e}x_{\rm e}$) in cm$^{-1}$, energy error with respect to CCSD(T) at the CCSD(T) equilibrium geometry ($\epsilon$) in m\Eh.
Results for C$_2$ and F$_2$ were included in the statistics only when available.
All results were computed using restricted Hartree--Fock orbitals and  freezing the Li--F 1s-like molecular orbitals.}
\label{tab:diatomics_errors}
\begin{tabular*}{6.5in}{@{\extracolsep{\fill}}cr@{}r@{}r@{}r@{}r@{}r@{}r@{}r@{}r@{}}
\hline\hline
Property & SRG(2) & SRG(2*) & \dsrg & \dsrg & \dsrg  & \dsrgph & \dsrgph & \dsrgph & CCSD \\
 & \multicolumn{1}{c}{$s$ = 5} & \multicolumn{1}{c}{$s$ = 5}  & \multicolumn{1}{c}{$s$ = 1}  & \multicolumn{1}{c}{$s$ = 10} & \multicolumn{1}{c}{$s$ = $\infty$} & \multicolumn{1}{c}{$s$ = 1} & \multicolumn{1}{c}{$s$ = 10} & \multicolumn{1}{c}{$s$ = $\infty$}  & \\
 \hline
$r_{\rm e}$\\
Mean & 0.0005 & -0.0052 & 0.0006 & 0.0004 & 0.0005 & -0.0088 & -0.0067 & -0.0067 & -0.0064 \\
Abs. Mean & 0.0016 & 0.0043 & 0.0063 & 0.0016 & 0.0015 & 0.0074 & 0.0056 & 0.0055 & 0.0053 \\
Std. Dev. & 0.0027 & 0.0053 & 0.0138 & 0.0027 & 0.0025 & 0.0089 & 0.0072 & 0.0072 & 0.0067 \\
Maximum & 0.0065 & 0.0163 & 0.0398 & 0.0068 & 0.0059 & 0.0268 & 0.0230 & 0.0230 & 0.0211 \\
$\omega_{\rm e}$\\
Mean & -11.9 & 40.0 & 10.4 & -13.1 & -13.9 & 67.9 & 45.9 & 45.6 & 43.8 \\
Abs. Mean & 17.1 & 33.4 & 38.6 & 18.7 & 18.4 & 56.5 & 38.3 & 38.0 & 36.5 \\
Std. Dev. &  23.0 & 37.2 & 66.7 & 25.2 & 23.8 & 50.5 & 40.0 & 39.8 & 39.0 \\
Maximum & 41.5 & 96.4 & 156.3 & 45.3 & 43.7 & 145.8 & 102.1 & 102.1 & 101.0 \\
$\omega_{\rm e} x_{\rm e}$\\
Mean & 1.1 & -0.4 & -2.5 & 1.7 & 1.7 & -2.2 & -0.3 & -0.4 & -0.3 \\
Abs. Mean & 1.0 & 0.4 & 2.3 & 1.4 & 1.5 & 1.9 & 0.4 & 0.4 & 0.4 \\
Std. Dev. & 1.3 & 0.5 & 3.9 & 1.6 & 1.6 & 2.5 & 1.0 & 1.0 & 0.8 \\
Maximum & 4.5 & 1.3 & 13.3 & 4.9 & 4.9 & 6.8 & 3.2 & 3.2 & 2.4 \\
$\epsilon$\\
Mean & -0.4 & 10.0 & 6.4 & -1.4 & -1.4 & 17.6 & 13.6 & 11.9 & 13.1 \\
Abs. Mean & 3.0 & 9.2 & 6.4 & 3.8 & 3.8 & 16.1 & 12.5 & 10.9 & 12.0 \\
Std. Dev. & 4.3 & 8.4 & 9.8 & 6.4 & 6.5 & 15.5 & 10.5 & 8.2 & 9.8 \\
Maximum & 9.7 & 19.3 & 35.2 & 17.9 & 18.4 & 60.8 & 36.9 & 20.9 & 33.6 \\
\hline\hline
\end{tabular*}
\end{table*}
\endgroup

A few interesting results are worth pointing out.
In several cases, the flow of the SRG(2) equations stalls and cannot be propagated after certain values.  Therefore, we present SRG(2) results for $s = 5$ $E_{\rm h}^{-2}$, which permits to converge most of the diatomic molecules (see below).
In the case of the DSRG(2), we report properties computed at $s$ = 1, 10, and $\infty$, observing that results from computations at $s$ = 10 $E_{\rm h}^{-2}$ agree well with the results obtained in the limit $s\rightarrow\infty$.
As found in the previous section, we notice that all properties computed with the SRG(2) and \dsrg are much closer to the CCSD(T) results than to the CCSD ones.
For example, the mean of the absolute deviations for $r_{\rm e}$ is ca. 0.0015 \AA{} both for the SRG(2) and \dsrg methods, while for CCSD the corresponding value is higher, 0.0053 \AA{}.

In addition, we encountered two problematic cases in which the SRG(2) and DSRG(2) methods fail.
The first one is C$_2$, for which neither the SRG(2) nor the \dsrg can be converged for $s \rightarrow \infty$.  We found that for smaller values of $s$ (0.1, 1, 2.5 $E_{\rm h}^{-2}$) it is possible to converge the \dsrg energy of C$_2$, but integration of the SRG(2) is not possible beyond $s$ = 0.0046 $E_{\rm h}^{-2}$, most likely because of the stiffness of the SRG equations.
The second problematic molecule is F$_2$.  In this case it is possible to find a bound minimum for both the SRG(2) and the \dsrg when $s$ is less than 1 $E_{\rm h}^{-2}$, but the resulting properties display unusually large errors.  
For example, with $s$ =1, the SRG(2) yields a value of $\omega_{\rm e} x_{\rm e}$ that is off by about 111 cm$^{-1}$.
Interestingly, for higher values of $s$ both the SRG(2) and the \dsrg appear to predict that F$_2$ is unbound.\bibnote{This result is inferred from the the observation that the SRG(2) and the \dsrg energy decreases as the F--F bond is stretched beyond the CCSD(T) equilibrium value, and that past a certain bond length, the energy cannot be converged.}
These two problems are reminiscent of some other notorious failures of approximated electron correlation methods.\cite{Pabst:2010fq}

Surprisingly, the (apparent) good agreement of the SRG(2), DSRG(2), and CCSD(T) results, as well as the failure of the SRG methods to describe C$_2$ and N$_2$ are not coincidental, but a consequence of the (2) truncation scheme.
As pointed out by us in a previous work,\cite{Evangelista:2011hz} when the commutator approximation $C=[A,B]_{1,2}$ is applied to the CCSD method, some third-order amplitude diagrams arising from the term $\frac{1}{2} [[V,T_2],T_2]$ have a weight that is half the correct value.
From this analysis it was suggested to correct the balance between third-order terms by doubling the one-body contributions to the commutator $[V,T_2]$ arising from the hole-hole and particle-particle diagrams. 
This modified commutator approximation permits to regain the correct balance between third-order terms in CCSD, and at the same it shifts the error to the fourth-order term $ [[V,T_1],T_2]$.

Here we show that a similar approximation that corrects the balance between third-order diagrams, here indicated with (2*), is beneficial to both the SRG and DSRG, and that in the case of the DSRG approach it also eliminates the problems encountered with C$_2$ and F$_2$.
The (2*) truncation is defined analogously to the (2) approximation, except that the following matrix elements are multiplied by a factor two:
\begin{align}
\tens{[A_1,B_2]}{i}{j} =& 2 \sum_{k}\sum_{c} (\tens{A}{k}{c} \tens{B}{ic}{jk} -\tens{A}{c}{k} \tens{B}{ik}{jc}), \\
\tens{[A_1,B_2]}{a}{b} =& 2 \sum_{k}\sum_{c} (\tens{A}{k}{c} \tens{B}{ac}{bk} -\tens{A}{c}{k} \tens{B}{ak}{bc}).
\end{align}
Error metrics for the diatomics computed using the commutator expansion defined by the (2*) approximation are also reported in Tables~\ref{tab:diatomics} and \ref{tab:diatomics_errors}.

Two points deserve attention.  First, errors for the SRG(2*) and \dsrgph methods computed with respect to CCSD(T) deteriorate, but become almost identical to those of the CCSD method.  This result indicates that as the commutator approximation is improved, the RG methods truncated to one- and two-body operators mimic the CCSD wave function.
This is in line with formal analyses\cite{Kutzelnigg:1991vl,Kutzelnigg:1998uy,Kutzelnigg:2010uv} and the empirical observation\cite{Cooper:2010ck,Evangelista:2011hz} of a \textit{no free lunch} result for single-reference electron correlation theories.
That is, in the case of a single-reference wave function, all electron correlation methods truncated to a certain particle rank (like for example, singles and doubles) represent a wave function with more or less the same degree of accuracy of coupled cluster theory.
Second, the modified commutator expansion improves the robustness of the \dsrg method.  This can be seen from the fact that \dsrgph computations on C$_2$ converge for all values of $s$ tested, and that this scheme predict the existence of a minimum for F$_2$.
On the contrary, C$_2$ and F$_2$ cannot be converged even with the improved SRG(2*) approximation.
In conclusion, although the SRG(2) and \dsrg methods show good agreement with the CCSD(T) results, this appears to be a consequence of error cancellation in the (2) approximation.
In view of these facts, (2*) is superior to the (2) approximation.

\subsection{Bond breaking with the DSRG approach}
\begin{figure}[ht]
\begin{center}
\includegraphics[width=3.5in]{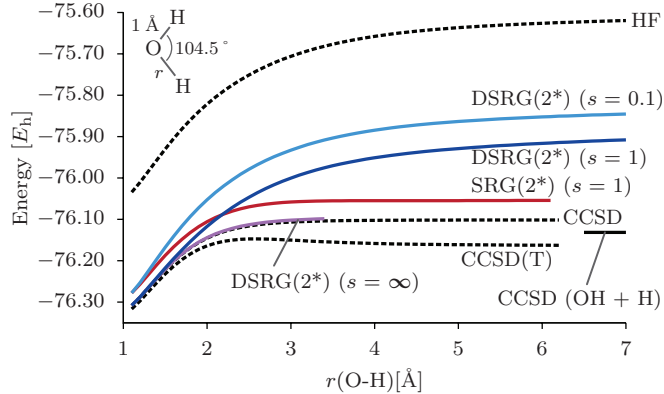}
\caption{Ground electronic state potential energy curve of water as a function of one of the O-H bond lengths.   The other O-H bond and the H-O-H angle were fixed to 1 \AA{} and 104.5 degrees, respectively.  Results for the SRG(2*) using the  White generator and the DSRG(2*) were based on the modified recursive approximation for the similarity-transformed Hamiltonian.  The figure also reports the energy computed at the Hartree--Fock, CCSD, and CCSD(T) levels of theory.  The sum of the CCSD energy for the OH and H fragments computed separately is indicated with a thick black line.  All results were computed using restricted Hartree--Fock orbitals and the cc-pVTZ basis set,  freezing the oxygen 1s-like molecular orbital.}
\label{fig:oh-h}
\end{center}
\end{figure}
In this section we consider the dissociation of an O--H bond in water as a test for the robustness of the SRG and DSRG methods.
It is important to reiterate that the main goal of the single-reference DSRG presented here is to exclusively describe dynamic electron correlation.
Since the DSRG does not try to diagonalize degenerate- or near-degnerate blocks of the Hamiltonian, in order to capture static correlation it is mandatory to formulate a multireference version of the DSRG in which the transformation of the Hamiltonian is coupled with the diagonalization of $\bar{H}$.

Fig.~\ref{fig:oh-h} shows the potential energy curve of H$_2$O computed with various many-body methods. 
At short bond distances, the CCSD and CCSD(T) curves appear to be well behaved, however, these methods cannot be converged past 6.2 \AA{}.
Both the CCSD and CCSD(T) methods do not converge to the proper dissociation limit (H + OH), which was computed as the sum of individual fragment energies at the CCSD level.  The CCSD(T) dissociation limit, which is not plotted in Fig.~\ref{fig:oh-h}, is almost indistinguishable from the reported CCSD value.
The SRG(2) (White generator, $s$ = 1 $E_{\rm h}^{-2}$) energy curve parallels the CCSD curve, recovering a smaller percentage of the correlation energy because the flow is not fully converged.
Like CCSD, the SRG(2*) curve is numerically stable up to about 6 \AA{}.

The DSRG(2*) curves computed for $s$ = 0.1 and 1 $E_{\rm h}^{-2}$ are well behaved converging in the entire range $r \in [1,7]$ \AA{}.  As $s$ is increased, the DSRG(2*) method recovers a larger amount of correlation energy, and in the limit $s\rightarrow\infty$ the DSRG(2*) follows the CCSD energy curve (although it is no longer numerically stable past ca. 3.4 \AA{}).
It is interesting to note that as the OH bonds is elongated, the DSRG(2*) curves with $s$ = 0.1 and 1 $E_{\rm h}^{-2}$ are found to parallel the Hartree--Fock energy curve.
To properly describe the dissociation of H$_2$O into H + OH, a zeroth-order wave function is required that contains three electronic configurations, generated by distributing two electrons in the O--H bonding  $(\sigma_{\rm OH})$ and antibonding orbitals $(\sigma^*_{\rm OH})$.
In this example, as the $(\sigma_{\rm OH})$ and $(\sigma^*_{\rm OH})$ orbitals become degenerate, the DSRG suppresses the amplitude corresponding to the excitations among these orbitals and cannot recover the proper dissociation limit.
Hence, the DSRG energy can be approximately divided into a dominant contribution from the Hartree--Fock reference determinant, which is responsible for the incorrect dissociation limit, plus a correction due to the dynamic component of electron correlation, which shifts the reference energy curve down.

\section{Summary and Discussion}
In this work we considered potential applications of similarity renormalization group (SRG) approaches to quantum chemistry.
Contrary to effective Hamiltonian theories that treat all excitations out of the reference on equal footing, in the SRG the flow equations are organized in such a way as to progressively zero certain elements of the Hamiltonian, starting with those that coupled the reference to the high-energy degrees of freedom.
The original formulation of SRG theory is based on a differential formalism:  the flow equations, a set of ordinary differential equations expressed in terms of a timelike variable $s$, must be propagated to compute the energy and other properties.
In the limit $s \rightarrow \infty$, the SRG flow performs a block diagonalization of the non-degenerate components of the Hamiltonian.  In this regard, the SRG is similar to other non-perturbative many-body approaches like the coupled cluster method, canonical diagonalization, and the anti-Hermitian Schr{\"o}dinger equation.
However, if the SRG flow is terminated at finite $s$, then this method permits to integrate out weak (dynamic) electron correlation from strong (static) correlation effects, yielding an effective renormalized Hamiltonian.
Because of this property, RG theories are attractive candidates for creating multireference many-body approaches that are numerically robust.  This was our motivation to begin the present study of the similarity renormalization group approach.

In order to address some shortcomings of the SRG, we introduce a novel \textit{integral} formulation of the SRG, which we refer to as the driven SRG (DSRG).
The DSRG transformed Hamiltonian, $\bar{H}(s)$, is parameterized by an exponential similarity transformation, which is generated by an excitation or anti-Hermitian operator $S(s)$.
In the DSRG, certain components of the similarity-transformed Hamiltonian are driven to zero by a source operator.
The source operator is required to satisfy boundary conditions compatible with the demand that $\bar{H}(s=0)$ coincides with the bare Hamiltonian, while $\bar{H}(s=\infty)$ is a block diagonal operator in which the reference Slater determinant is decoupled from all the excited determinants.
The DSRG is a full fledged many-body renormalization approach and it is computationally more  advantageous than the SRG since it consists of a set of coupled polynomial equations.  Furthermore, in the single-reference case, the DSRG equations may be viewed as modified coupled cluster or unitary coupled cluster equations.

We present the first comprehensive study of the ground state equilibrium properties of a number of diatomic molecules using the SRG method and the unitary variant of the DSRG approach.
In particular, we consider the case of a single-reference vacuum, and use the (2) truncation scheme, which is defined by writing all operators in normal ordered form and neglecting three- and higher-body operators.
It is found that for finite and sufficiently large values of the timelike parameter $s$, these two methods yield results that are closer to the CCSD(T) method.   
However, we discovered a serious failure of the (2) approximation: the SRG(2) and DSRG(2) equations cannot be converged in the case of C$_2$, and both methods appear to predict that F$_2$ is unbound in the limit $s \rightarrow \infty$.
A modified approximation, (2*), correct up to fourth-order, fixes the deficiencies of the (2) approximation in the case of the DSRG(2*) approach and yields results that are in close agreement with the CCSD ones.
While there is certainly room for improving the truncated BCH series,\cite{Neuscamman:2009cya} the (2*) approximation appears to be a viable approach that deserves further scrutiny.

The theoretical developments and numerical results presented here form the basis for exploring a number of applications of similarity renormalization group ideas to electronic structure theory.
The most interesting aspect to study is the generalization of the DSRG to a multiconfigurational reference wave function.
In this regard, the DSRG is expected to provide a simple solution to the problem of intruders that arises in multireference coupled cluster theories.
However, there are a number of other intriguing problems that can be addressed with the DSRG approach.
Recent work on spin-component scaled and attenuated perturbation theories\cite{Grimme:2003va,Jung:2004bp,Goldey:2013hs} suggests that the variable $s$ might be considered as a parameter to fine tune the accuracy of the DSRG second-order perturbation theory and DSRG(2) methods.
The DSRG might also be used to address the electronic structure of extended metallic systems for which zero- and finite-temperature second-order perturbation theories are known to diverge.\cite{Hirata:2013bp,He:2014bg}
Last, because the DSRG induces a renormalization of all interactions above the energy cutoff $\Lambda$, it also provides an interesting framework for inventing new adaptive quantum chemistry methods.\cite{Bender:1969va,Buenker:1974bl,Huron:1973cb,Nakatsuji:1991jt,Abrams:2005ui,Bytautas:2009hm,Lyakh:2010ke,Thom:2010ua,Landau:2010hz,Melnichuk:2012kr,Tenno:2013kd,Evangelista:2014ko}

\begin{acknowledgments}
The author is grateful to Philip Shushkov for insightful discussions on renormalization theory.  John F. Stanton is acknowledged for suggesting the water potential energy curve as an example of system for which CCSD fails to converge.  This work was supported by start-up funds provided by Emory University.
\end{acknowledgments}

\appendix*
\section{Matrix elements of $C=[A,B]_{1,2}$} \label{app:matrix_elements}
An operator $A$ containing at most two-body terms may be written in normal ordered form with respect to the reference $\mref$ as:
\begin{equation}
A = A_0 + A_1 + A_2,
\end{equation}
where $A_0$ is a scalar, and
\begin{align}
A_1 &= \sum_{pq} \tens{A}{p}{q} \no{\sqop{p}{q}},\\
A_2 &= \sum_{pqrs} \tens{A}{pq}{rs} \no{\sqop{pq}{rs}},
\end{align}
with the second quantization operator written compactly as $\sqop{p}{q} = \cop{p} \aop{q}$ and $\sqop{pq}{rs} = \cop{p} \cop{q} \aop{s} \aop{r}$.
The commutator $C=[A,B]_{1,2}$ contains contributions from the following terms:
\begin{align}
C_0 &= \bra{\mref}[A_1,B_1]\ket{\mref} + \bra{\mref}[A_2,B_2]\ket{\mref}, \\
\tens{C}{p}{q} &= \tens{[A_1,B_1]}{p}{q}  + \tens{[A_1,B-side]}{p}{q} 
- \tens{[B_1,A_2]}{p}{q} + \tens{[A_2,B_2]}{p}{q},\\
\tens{C}{pq}{rs} &= \tens{[A_1,B_2]}{pq}{rs} - \tens{[B_1,A_2]}{pq}{rs} + \tens{[A_2,B_2]}{pq}{rs},
\end{align}
where the unique contributions to the matrix elements are:
\begin{align}
\bra{\mref}[A_1,B_1]\ket{\mref} =& \sum_{p} \sum_i (\tens{A}{i}{p} \tens{B}{p}{i}- \tens{B}{i}{p} \tens{A}{p}{i}), \\ \displaybreak[0]
\bra{\mref}[A_2,B_2]\ket{\mref} =& \frac{1}{4} \sum_{ij} \sum_{ab} (\tens{A}{ij}{ab} \tens{B}{ab}{ij}-\tens{B}{ij}{ab} \tens{A}{ab}{ij} ),
\end{align}
\begin{align}
\tens{[A_1,B_1]}{p}{q} =& \sum_r (\tens{A}{p}{r} \tens{B}{r}{q} - \tens{B}{p}{r}\tens{A}{r}{q}),\\ \displaybreak[0]
\tens{[A_1,B_2]}{p}{q} =& \sum_{i}\sum_{a} \tens{A}{i}{a} \tens{B}{pa}{qi} -\tens{A}{a}{i} \tens{B}{pi}{qa},\\  \displaybreak[0]
\tens{[A_2,B_2]}{p}{q} =&\frac{1}{2}\sum_{ij}\sum_{a} \left( 
\tens{A}{ap}{ij}\tens{B}{ij}{aq}
-\tens{A}{ij}{aq}\tens{B}{ap}{ij}
\right)\\
&+\frac{1}{2}\sum_{i}\sum_{ab}
\left( 
\tens{A}{ip}{ab}\tens{B}{ab}{iq}
-\tens{A}{ab}{iq}\tens{B}{ip}{ab}
\right), \nonumber
\end{align}
\begin{align}
\tens{[A_1,B_2]}{pq}{rs} =& \sum_t \left[P(pq) \tens{A}{p}{t} \tens{B}{tq}{rs} - P(rs) \tens{A}{t}{r} \tens{B}{pq}{ts}\right],\\
\tens{[A_2,B_2]}{pq}{rs} =&
\frac{1}{2}\sum_{ab}
(\tens{A}{pq}{ab}\tens{B}{ab}{rs} -\tens{A}{ab}{rs} \tens{B}{pq}{ab})\\
&-\frac{1}{2}\sum_{ij}
(\tens{A}{pq}{ij}\tens{B}{ij}{rs} -\tens{A}{ij}{rs} \tens{B}{pq}{ij})\nonumber\\
&+\sum_{i}\sum_{a}
P(pq) P(rs) \left[\tens{A}{pi}{ra}\tens{B}{qa}{si} -
 \tens{A}{pa}{ri}\tens{B}{qi}{sa} \right].\nonumber
\end{align}


\begin{thebibliography}{128}
\bibitem [{1()}]{1}%
  \BibitemOpen
  \href@noop {} {}\bibinfo {note} {Notice that in the notation used here the
  top and bottom indices of the matrix elements of the many-body operators are
  switched with respect to the traditional notation. For example, the coupled
  cluster amplitudes are written as $\tens {t}{ab}{ij}$ instead of $\tens
  {t}{ij}{ab}$. However, this way of writing indices reflects the proper tensor
  character of the many-body operators.}\BibitemShut {Stop}%
\bibitem [{2()}]{2}%
  \BibitemOpen
  \href@noop {} {}\bibinfo {note} {The matrix elements $\tens {\eta }{i}{a}(s)$
  and $\tens {\eta }{ij}{ab}(s)$ are computed from Eqs.~\eqref
  {eq:white_generator_s} and \eqref {eq:white_generator_d} as $\tens {\eta
  }{i}{a}(s) = -\tens {\eta }{a}{i}(s)$ and $\tens {\eta }{ij}{ab}(s) = -\tens
  {\eta }{ab}{ij}(s)$.}\BibitemShut {Stop}%
\bibitem [{3()}]{3}%
  \BibitemOpen
  \href@noop {} {}\bibinfo {note} {This problem is analogous to the one
  encountered in unitary coupled cluster theory, where it is necessary to
  truncate the Baker--Campbell--Hausdorff expansion for $\bar {H}$. However, in
  the coupled cluster version of DSRG, the similarity-transformed Hamiltonian
  truncates after four commutators, and thus, it is not necessary to
  approximate the corresponding amplitude equations.}\BibitemShut {Stop}%
\bibitem [{4()}]{4}%
  \BibitemOpen
  \href@noop {} {}\bibinfo {note} {First-order perturbation theory does not
  provide insight into the form of the triples and higher amplitude equations,
  and several options are possible. One possibility is a generalization of the
  singles and doubles equations for $R(s)$: \begin {equation} \tens
  {r}{ab\cdots }{ij\cdots }(s) = \, \left [\tens {\bar {H}}{ab\cdots }{ij\cdots
  }(s) + \tens {t}{ab\cdots }{ij\cdots } \denominator {ab\cdots }{ij\cdots
  }\right ] \, e^{-s (\denominator {ab\cdots }{ij\cdots })^2}, \end {equation}
  but the simpler condition: \begin {equation} \tens {r}{ab\cdots }{ij\cdots
  }(s) = 0, \end {equation} also appears to be viable.}\BibitemShut {Stop}%
\bibitem [{5()}]{5}%
  \BibitemOpen
  \href@noop {} {}\bibinfo {note} {This result is inferred from the the
  observation that the SRG(2) and the \dsrg energy decreases as the F--F bond
  is stretched beyond the CCSD(T) equilibrium value, and that past a certain
  bond length, the energy cannot be converged.}\BibitemShut {Stop}%
\end{thebibliography}
%
\end{document}